\documentclass[structabstract]{aa} 
\usepackage{graphicx}
\usepackage{upgreek}
\usepackage{amsmath}

\usepackage{natbib}
\usepackage{twoopt}
\usepackage{multirow}
\usepackage{caption}
\usepackage{siunitx}
\usepackage{gensymb}
\usepackage{threeparttablex}
\usepackage{subfig}

\usepackage{mathtools}
\usepackage{braket}

\usepackage{txfonts}
\usepackage{color}
\usepackage{hyperref}
\hypersetup{%
  colorlinks=true,
  linkcolor=blue,
  citecolor=blue
}

\def \bmath #1 {{\hbox{\boldmath{$#1$}\unboldmath}}}

\usepackage[dvipsnames]{xcolor}

\begin{document}

\title{Inelastic $\mathrm{H} + \mathrm{H}_3^+$ Collision rates and their impact in the determination of the excitation temperature of $\mathrm{H}_3^+$}
\authorrunning{F\'elix-Gonz\'alez et al.}
\author{Daniel F\'elix-Gonz\'alez\inst{1}
  \and
  Pablo del Mazo-Sevillano\inst{1}
  \and
  Alfredo Aguado\inst{1}
  \and
  Octavio Roncero\inst{2}
  \and
  Jacques Le Bourlot\inst{3,4}
  \and
  Evelyne Roueff\inst{3}
  \and
  Franck Le Petit\inst{3}
  \and
  Emeric Bron\inst{3}
}
\institute{
  Unidad Asociada UAM-IFF-CSIC, Departamento de  Qu{\'\i}mica F{\'\i}sica Aplicada, Facultad de Ciencias M-14, Universidad Aut\'onoma de Madrid, 28049 Madrid (Spain)       
          \and
          Instituto de F{\'\i}sica Fundamental (IFF-CSIC), C.S.I.C., Serrano 123, 28006 Madrid, Spain. \email{octavio.roncero@csic.es}
          \and
          Sorbonne Universit\'e, Observatoire de Paris, Univerit\'e PSL, CNRS, LERMA, 92190 Meudon, France
          \and
          Université Paris-Cité
}
\titlerunning{H+H$_3^+$ collision rates}
\abstract
{
  In diffuse interstellar clouds the excitation temperature derived
  from the lowest levels of $\mathrm{H}_3^+$ is systematically lower
  than that derived from $\mathrm{H}_2$.  The differences { may be} attributed to the lack of state-specific  formation and destruction rates  of $\mathrm{H}_3^+$ needed to thermalize the two species.
}
{
In this work, we want to check the role of rotational excitation collisions of $\mathrm{H}_3^+$ with atomic hydrogen on its excitation temperature.
}
{A time independent close-coupling method is used to calculate the state-to-state rate coefficients,
  using a very accurate and full dimensional potential energy surface recently developed for H$_4^+$. A symmetric top approach is used to describe a frozen H$_3^+$ as equilateral triangle. 
}
{
Rotational excitation collision rate coefficients of $\mathrm{H}_3^+$ with atomic Hydrogen have been derived in a temperature range appropriate to diffuse interstellar conditions up to $(J,K,\pm)= (7,6,+)$ and $(J,K,\pm) = (6,4,+)$ for its { ortho} and { para} forms. This allows to have a consistent set of collisional excitation rate coefficients and to improve the previous study where these contributions were speculated.
}
{
The new state-specific inelastic $\mathrm{H}_3^+ + \mathrm{H}$ rate coefficients yield differences up to $20\,\%$ in the excitation temperature, and their impact increases with decreasing molecular fraction.
We also confirm the impact of chemical state-to-state destruction reactions in the excitation balance of $\mathrm{H}_3^+$,
and that reactive $\mathrm{H} + \mathrm{H}_3^+$ collisions are also needed to account for possible further { ortho} to { para} transitions.
}

   \keywords{Astrochemistry -- ISM: abundances -- ISM: molecules }

  \maketitle

\section{Introduction}

Hydrogen is the most abundant element in Universe and plays a key role in characterizing the physical conditions of the interstellar medium (ISM),  the star formation and the chemical evolution of the molecular Universe \citep{Oka:13}. Among its molecular forms ($\mathrm{H}_2$, $\mathrm{H}_2^+$ and $\mathrm{H}_3^+$),  $\mathrm{H}_3^+$ is an efficient protonator as the proton affinity of $\mathrm{H}_2$ ($422.3 \, \mathrm{kJ/mol}$) is smaller than that of most stable molecules \citep{Watson:73,Herbst-Klemperer:73,Millar-etal:89,Pagani-etal:92,Tennyson:95,McCall-Oka:00,Oka:12}. For an atom or molecule (generally $\mathrm{M}$) the reaction is: 
\begin{eqnarray}
  \mathrm{M} + \mathrm{H}_3^+ &\rightarrow& \mathrm{MH}^+ + \mathrm{H}_2
\end{eqnarray}
The ionic hydrides thus formed trigger the chemistry cycles of many complex molecules in space \citep{Watson:73,Herbst-Klemperer:73}. 

$\mathrm{H}_3^+$  has been the subject of many review studies \citep{Tennyson:95,Herbst:00,Oka:12,Gerlich-etal:12,Oka:13,Miller-etal:20}
and special issues \citep{H3+special-issue:12,H3+special-issue:19}.
Its infrared spectrum was first  detected in  the laboratory by \citet{Oka:80} and later
in space \citep{Geballe-Oka:89,Geballe-etal:99,McCall-etal:99,Oka:13}. Since then, $\mathrm{H}_3^+$ has been used to probe
spatial conditions, as a thermometer and a clock of cold molecular clouds \citep{Oka:06,Pagani2011} and as a measure of the ionization rate of ISM \citep{2004A&A...417..993L,2012ApJ...745...91I} and in the Central Molecular Zone of our Galaxy \citep{2016A&A...585A.105L}. Its infrared
spectrum has been theoretically characterized with spectroscopic accuracy
\citep{Polyanski2012:5014,Bachorz2009:024105,Velilla2008:084307,Tennyson2017:232001,Furtenbacher2013:5471},
based on highly accurate
potential energy surfaces (PESs) 
\citep{Jaquet-etal:98b,Tennyson:95,Cencek1998:2831,Pavanello2012:184303,Mizus2018:1663,Bachorz2009:024105,Velilla2010:387,Rohse1994:2231,Viegas2007:074309,Ghosh2017:074105}.

 The absence of permanent dipole moment in the highly symmetric triangular equilibrium geometry  in its ground vibrational state makes $\mathrm{H}_3^+$ unobservable using pure rotational spectroscopy.
 $\mathrm{H}_3^+$ is thus only observable through its vibrational spectrum. The recent availability of high sensitivity infrared observations thanks to the James Webb Spatial Telescope (JWST) has even allowed to detect infra red emission of $\mathrm{H}_3^+$ in ultra luminous infra red galaxies (ULIRGs) \citep{Pereira-Santaella-etal:24}.
 This opens the opportunity of a wider use of $\mathrm{H}_3^+$ as a probe of the physical conditions of different objects in ISM.

 The $\mathrm{H}_3^+ + \mathrm{H}_2$ collisional rates, including { ortho/para} transitions of the two species, have been reported in several theoretical studies \citep{Oka-Epp:04,Hugo-etal:09,Park-Light:07,Gomez-Carrasco-etal:12}. However the collisional rates with atomic hydrogen are yet non available despite their potential importance in partially molecular environments such as diffuse and translucent clouds and in the Central Molecular Zone of our Galaxy
 \citep{Miller-etal:20,Oka:12}. 
 
 The deuteration rates in $\mathrm{D} + \mathrm{H}_3^+$ and isotopic variants reactive collisions
 have been studied experimentally \citep{Hillenbrand-etal:19,Bowen-etal:21}. Due to the vibrational excitation in which $\mathrm{H}_3^+$ is initially formed,
 a pure experimental determination of the deuteration rates is not possible. For this reason,
 a combined experimental and theoretical treatment has been used to determine the reactive deuteration rate constants.
 In such studies, tunneling through a potential barrier was estimated theoretically,
 but produced rate constants lower than $10^{-12} \, \mathrm{cm}^3 \, \mathrm{s}^{-1}$  below $100 \, \mathrm{K}$.
 Ring Polymer Molecular Dynamics calculations \citep{Bulut-etal:19},
 including quantum effects, on these reactions agree pretty well with these combined experiment/theoretical results \citep{Hillenbrand-etal:19,Bowen-etal:21}
 above $200 \, \mathrm{K}$, but below this temperature there is no full consensus about the role of the tunneling rate. {  Using the tunnelling values
    reported by  Bowen {\it et al.} (2021) as an upper limit, we can conclude that
    the reactive exchange rate is lower than 10$^{-12}$ cm$^3$s$^{-1}$ below 100 K
    and 10$^{-10}$ cm$^3$s$^{-1}$ below 300 K. Under this circunstance H+H$_3^+$ rotational inelastic
    collisions can then be considered as the dominant process below 100 K.}

\begin{figure}
    \center
    \includegraphics[scale=0.35]{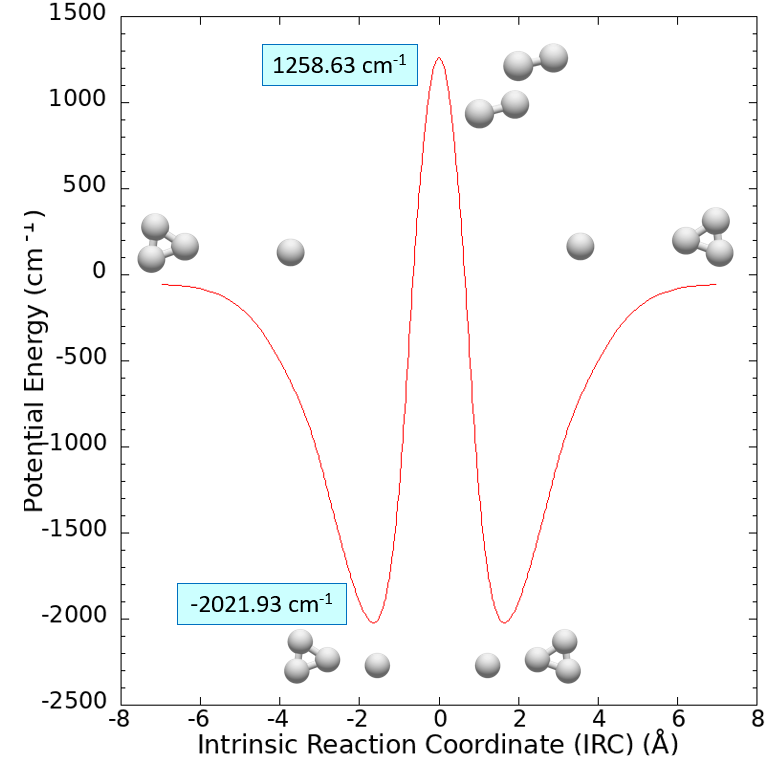}
    \caption{Minimum energy path for the H + H$_3^+$ $\longrightarrow$ H$_3^+$ + H exchange reaction for the PES by \cite{delMazo-Sevillano-etal:24a}}
    \label{fig:Profile}
\end{figure}

  The exchange barrier is of approximately $1258 \, \mathrm{cm}^{-1}$, as shown in Fig.~\ref{fig:Profile},
  according to a very accurate potential energy surface (PES) recently developed to study
  the $\mathrm{H}_2 + \mathrm{H}_2^+ \rightarrow \mathrm{H}_3^+ + \mathrm{H}$ reaction \citep{delMazo-Sevillano-etal:24a}.
  { The complete quantum treatment of inelastic
    and reactive dynamics, specially considering ortho/para permutation
    symmetry, is now-a-days inaffordable, and new methods need to be developed
    to address that problem. Before addressing the full exact treatment, in this work
    we study the   H+H$_3^+$ rotational inelastic collisions at low temperatures, corresponding majoritarily to energies below the top of the barrier, to determine
  its role in the excitation temperature of  $\mathrm{H}_3^+$.}

In diffuse interstellar clouds the excitation temperature $T_{12}(\mathrm{H}_3^+)$ derived from the lowest two levels of $\mathrm{H}_3^+$ is systematically lower than the gas kinetic temperature as derived from $\mathrm{H}_2$ lowest levels $T_{01}(\mathrm{H}_2)$. \citet[][hereafter Paper I]{2024MolPh.12282612L} give a thorough discussion of $\mathrm{H}_3^+$ excitation mechanisms and propose an explanation of this observational fact. $T_{12}(\mathrm{H}_3^+)$ is defined by:
\begin{equation}
    T_{12}(\mathrm{H}_3^+) = E_{12} \, / \, \ln\left({\frac{g_2 x_1}{g_1 x_2}}\right)
\end{equation}
Where $E_{12} = 32.86\, \mathrm{K}$ is the energy difference between both levels, $g_2/g_1 = 2$ and $x_1$ and $x_2$ are the populations of levels $1$ and $2$ respectively. This difference comes as a surprise as $\mathrm{H}_3^+$ is formed in the region where $\mathrm{H}_2$ dominates and the lowest levels of both species should be thermalized at the gas kinetic temperature. As in most lines of sight only these $2$ lowest levels are observed, all interpretation of $\mathrm{H}_3^+$ excitation rely on that single value.

Paper I shows that the difference arises from the fact that state-specific formation and destruction processes of $\mathrm{H}_3^+$ must be accounted for when computing the molecule detailed balance steady state. In essence, a significant fraction of the formation reaction $\mathrm{H}_2^+ + \mathrm{H}_2$ exothermicity populates high lying levels of $\mathrm{H}_3^+$ \citep{delMazo-Sevillano-etal:24a} favoring the {para} form at low temperature by nuclear spin selection rules \citep{Oka:04}. These levels decay efficiently by radiative transitions to the lowest accessible level, which is either the {para}-$(1,1,-)$ level or the {ortho}-$(3,3,-)$ level (which is metastable). But the lower lying {ortho}-$(1,0,+)$ level can only be populated by
{ slow  reactive collisions}, mainly with $\mathrm{H}_2$. This process comes into competition with destruction of the molecule by dissociative recombination with $e^{-}$, leading to an underpopulation of the lowest lying {ortho} level compared to a Boltzmann population at the gas kinetic temperature.

This mechanism is very sensitive to all the various state to state rates used. In the absence of better data, Paper I  approximates collisions with $\mathrm{H}$ by taking $\mathrm{H}_2$ rate coefficients scaled by a mass factor of $\sqrt{2}$. This was done for all transitions, including reactive ones. However, if reactive rate coefficients with $\mathrm{H}$ are negligible, this may lead to significant differences in a medium which is not fully molecular.
{ \cite{2024MolPh.12282612L}  also showed that} possible small differences in the dissociative recombination rates of {ortho} and {para} modifications may have a significant impact on the excitation temperature.

In the following, the inelastic cross sections and rate constants are presented and discussed in Section~\ref{sec:Results} and their impact on two typical examples of diffuse lines of sight is presented in Section~\ref{sec:Astro}.

\section{Inelastic Scattering Results}\label{sec:Results}

In this work we use the full dimensional PES developed by \citet{delMazo-Sevillano-etal:24a}, which considers all the degrees of freedom of the 4 atom systems very accurately using a Neural Network (NN) method \citep{delMazo-Sevillano-etal:24a}. In addition, this PES describes very well the long-range interaction through the use of a triatom-in-molecules (TRIM) formalism \citep{Sanz-Sanz-etal:13,Sanz-Sanz-etal:15}, originally proposed for $\mathrm{H}_5^+$ system \citep{Aguado-etal:10}.  

The $\mathrm{H} + \mathrm{H}_3^+$ inelastic rotational collisions are studied  below $1500 \, \mathrm{cm}^{-1}$ of translational energy, considering the rigid-rotor approach and the {ortho/para} symmetry of $\mathrm{H}_3^+$. To this aim we proceed in two steps.  First, the triatomic levels of $\mathrm{H}_3^+$ are calculated in full dimension, using permutationally invariant basis set functions represented in hyperspherical coordinates \citep{Aguado-etal:00,Sanz-etal:01}, as 
described in the Appendix~\ref{sec:H3+-levels}. Second, the inelastic rotational cross sections and rate constants are calculated using a close-coupling approach within a rigid rotor approach, using a symmetric top + atom adapted from  the study of NH$_3$ + He \citep{Green1980} and briefly 
described in Appendix~\ref{sec:symmetric-top-scattering}.

\begin{figure*}
\center
 \includegraphics[scale=0.4]{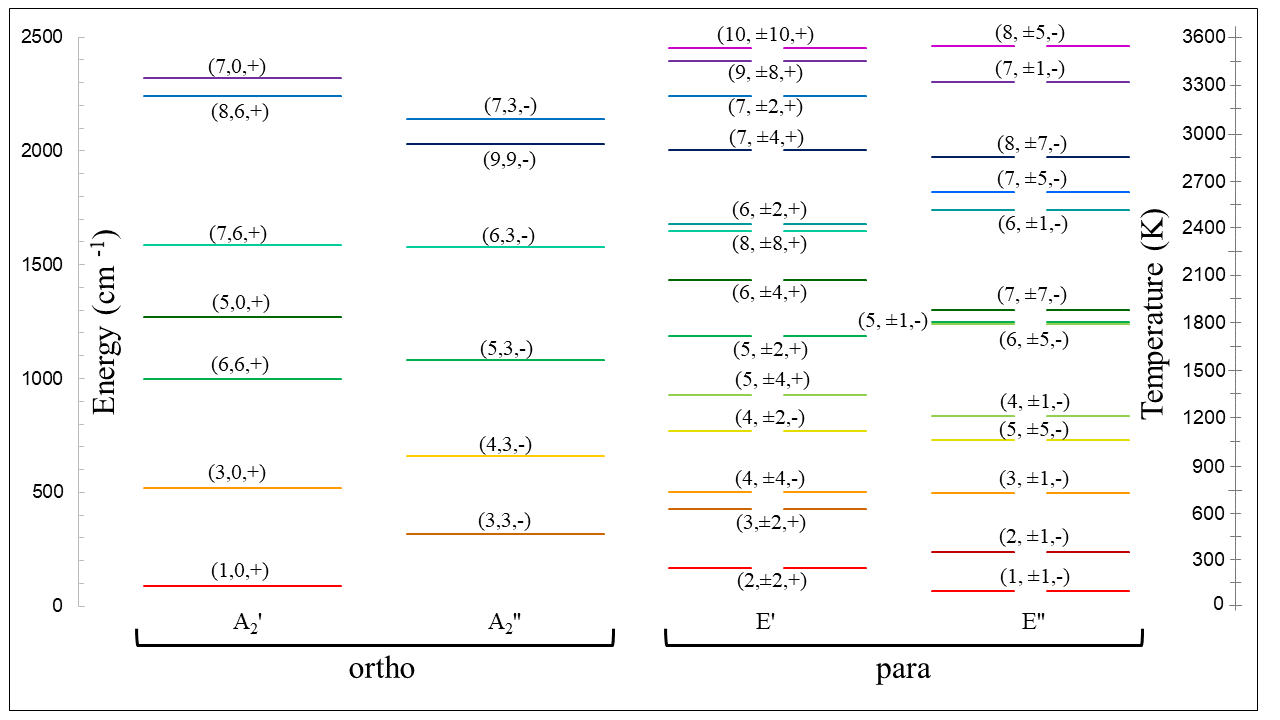}
 \caption{Computed H$_3^+$ rotational levels in the ground vibrational state, conventionally labelled as $(j,\omega,p_t)$ within each specific irreducible representation, $\Gamma$.}
 \label{fig:H3p-levels}
\end{figure*}

The resolution of the close-coupling equations is done as follows.
A radial grid of $5000$ equidistant points is used to describe $R$,  from $1 \,$ to $100 \,$ bohr. In view of the energy requirements discussed for the basis selection (triatomic basis up to $4100 \, \mathrm{cm}^{-1}$), in this study we have considered $4000$
equispaced { total  energies} from $70 \, \mathrm{cm}^{-1}$ to $2070 \, \mathrm{cm}^{-1}$, with a step of $0.5 \, \mathrm{cm}^{-1}$, measured from the unphysical ground $A_1, J=0$, $(0,0^0$) $\mathrm{H}_3^+$ level. The calculations are done up to a maximum total angular momentum $J=30$, and separately for $\Gamma_t = A_2$ ({ortho}-$\mathrm{H}_3^+$) and $E$ ({para}-$\mathrm{H}_3^+$) representations.
The $\mathrm{H}_3^+$ states finally considered are shown in Fig.~\ref{fig:H3p-levels}, for each irreducible representation $\Gamma$, and reducing the quantum numbers to ($j,\omega, p_t, \Gamma_t$), which are the only needed in the scattering calculations presented below (see Appendix~\ref{sec:H3+-levels}).
 
Let us recall that the channels $\alpha \equiv (\beta,\ell)$ naturally arise from the triatomic levels $\beta \equiv (j,\omega,p_t, \Gamma_t)$ by introducing in each case all the allowed values for $\ell = \vert J-j \vert,\ldots,{J + j}$. This implies that the number of channels rapidly increases with increasing the total angular momentum $J$. 
An additional truncation of channels is imposed. For triatomic levels above $j = 6$ ($1200 \, \mathrm{cm}^{-1}$) within the $\Gamma_t = E$ representation, the simulated channels are constrained to $\ell \le {J+j}/2$. This reduces the computation times, while the net effect on the final rate constants is small --- note that most truncated channels are either predominantly closed or less contributing to the averages.

Transitions involving lower  ${\Delta j}$, ${\Delta \omega}$, and $\Delta p_t = 0$  are expected to yield larger cross sections, simply because 
the radial potential coefficients, $V_{\lambda\nu}$ as defined in equation B.5, connecting them are larger
for smaller $\lambda$ and $\nu$ values, as can be seen in figure ~\ref{fig:Vlamnu}.

It is expected that transitions involving fewer changes in the internal state of the system are more favored --- {\it i.e.} higher $\sigma_{\beta\beta'}$ for lower ${\Delta j}$, ${\Delta \omega}$, and $\Delta p_t = 0$. 
From a mathematical point of view, the transition probability is maximized when the dominant radial coefficients $V_{\lambda\nu}$ in that transition are greater. By direct examination of figure \ref{fig:Vlamnu}, it is clear that the coefficients are bigger in magnitude for smaller values of $\lambda$ and $\nu$, the latter being more critical --- for example, $V_{\lambda = 2, \nu = 0}$ is comparable to $V_{\lambda = 3, \nu = 3}$. As long as $\nu = \Delta \omega$ and $\lambda \geq \Delta j$, the intuitive rules are completely equivalent to the rigorous mathematical criterion.

\begin{figure}
    \centering
    \includegraphics[scale=0.38]{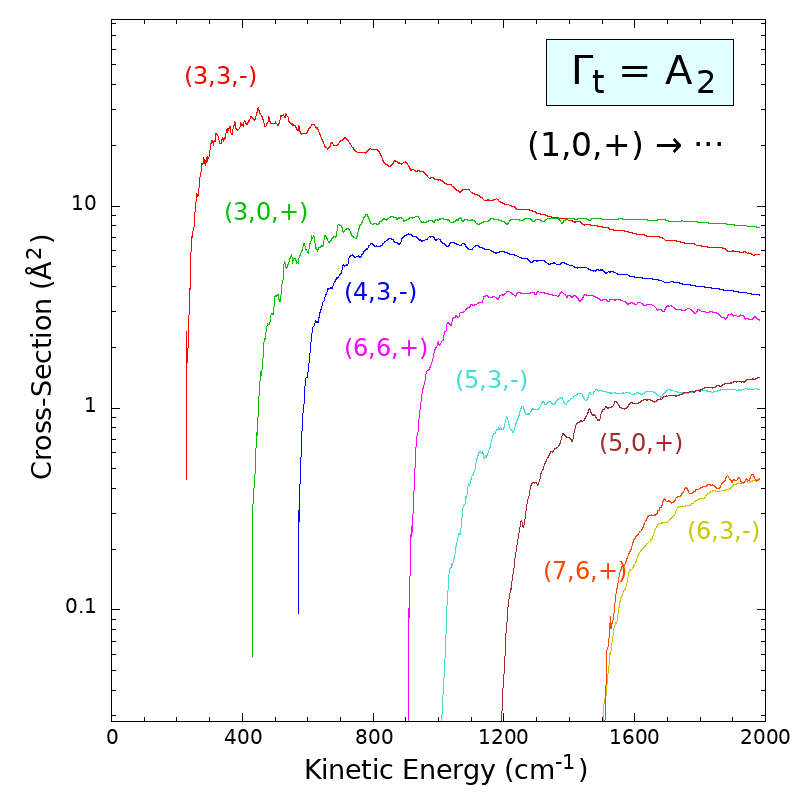}
    \caption{Calculated cross-sections for the transitions starting from (1,0,+) $\rightarrow (j',\omega',p_t')$ within the $\Gamma_t = A_2$ representation.}
    \label{fig:A2-sigma0001}
\end{figure}

Apart from this rule, it is important to consider that the channels have an energy threshold. The radial coefficients of the dissociative wave function
oscillate with a frequency proportional to $\sqrt{2\mu (E-E_{j\omega,p_t})}$. Having closer thresholds implies similar frequencies, what maximizes the effective radial integrals,  yielding larger transition probabilities.

For the $\Gamma_t = A_2$ representation ({ortho}-$\mathrm{H}_3^+$), we present in figure \ref{fig:A2-sigma0001} the transitions starting from $(1,0,+)$, the lowest level in this representation. In this representation,
the $\omega$ quantum number for all the levels are positive integers multiple of three, and the transitions correspond to  $\nu = \Delta \omega$ which are always multiples of three. In other words, all the transitions for this representation are symmetry allowed.

Starting from (1,0,$+$), the most favored transition would be to the (3,0,$+$) level,
as it is energetically close to the entrance level and fulfils $\Delta \omega = 0$, $\Delta j = 2$
[$V_{\lambda=2,\nu=0}$] and $\Delta p_t = 0$. In view of figure \ref{fig:A2-sigma0001}, this statement is apparently only true
for higher energies, but at lower energies the transitions to (3,3,$-$) dominate,
with $\Delta \omega = 3$, {  $\Delta j = 2$} [$V_{\lambda=3,\nu=3}$] and $\Delta p_t \neq 0$. 
The coefficients $V_{\lambda=2,\nu=0}$ and $V_{\lambda=3,\nu=3}$ have a comparable magnitude, but the (3,3,$-$)
level is closer in energy to the (1,0,$+$) entrance level. In such situation the energy criterion described above dominates.

A similar case occurs for the transitions to (5,0,$+$) [$V_{\lambda=4,\nu=0}$]. For energies below its threshold other less favoured transitions have a higher probability than expected (such as (5,3,$-$) [$V_{\lambda=5,\nu=3}$]). However, once the level opens, its transition probability rapidly increases and finally dominates over the former one.

\begin{figure}
    \centering
    \includegraphics[scale=0.38]{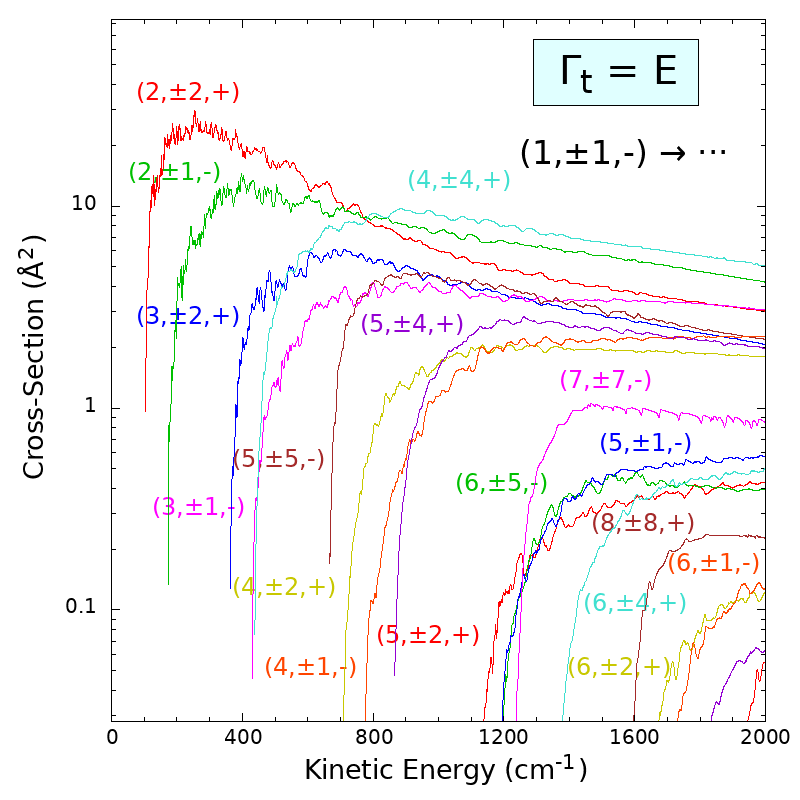}
    \caption{Calculated cross-sections for the transitions starting from (1,$\pm$1,-) within the $\Gamma_t = E$ representation.}
    \label{fig:E-sigma0001}
\end{figure}

For the $\Gamma_t = E$ representation ({para}-$\mathrm{H}_3^+$), the same propensity rules apply, but some slight details must be taken into consideration. This time, for each energy level there are two degenerate states associated to the $\pm \omega$ projections. For a generic transition between two levels $\beta$ and $\beta'$ (each containing two degenerate states $\pm \omega$ and $\pm \omega'$, respectively), four possible sub-transitions may occur. According to the selection rules, two of these sub-transitions are always forbidden, while the other two have  exactly the 
same cross-section. As a consequence, the final average associated to that transition is in practice equivalent to taking just one of the allowed sub-transitions.

Additionally, as already stated by \cite{Bouhafs2017} for analogous systems, the triatomic levels for this representation have values of $\omega$ which are not multiples of three. As a consequence, not all the transitions will have values of $\nu = \Delta \omega$ that are multiples of three, so there will be some symmetry-forbidden transitions.

In view of figure \ref{fig:E-sigma0001}, starting from $(1,\pm1,-)$, the most favored transition would be to the $(2,\pm 2, +)$ level because of its energetic proximity and with ${\Delta\omega} = 3, \Delta j = 1$ [$V_{\lambda = 3, \nu = 3}$],  $\Delta p_t \neq 0$. Then, once the $(2,\pm 1, -)$ level opens, its transition becomes dominant as expected --- ${\Delta\omega} = 0, \Delta j = 1 ~[V_{\lambda = 1, \nu = 0}], \Delta p_t = 0$. 

There is an interesting phenomenon associated to this symmetry representation,
which can be exemplified by the transitions to $(3,\pm 2 ,+)$ and $(4, \pm 4, +)$.
They both correspond to the element $V_{\lambda = 3, \nu = 3}$ --- ${\Delta\omega} = 3, \Delta p_t \ne 0$, with { $\Delta j = 2$ and $\Delta j = 3$}, respectively. Therefore, considering their relative energies, one may expect the former level to be more favored, but as we examine figure \ref{fig:E-sigma0001} we see that $(4, \pm 4, +)$ is indeed the dominant transition. This is because, within $\Gamma_t = E$, there is an extra propensity rule in relation to the sign of $\omega$. Transitions from $(1,\pm 1, -)$ to $(3,\pm 2 ,+)$ imply a change in the sign of $\omega$, while $(4, \pm 4, +)$ leaves the sign unaltered.

This physical intuitive argument can be algebraically justified if we take special attention to the expression of the 3-j symbols appearing in the angular elements of the potential, Eq.~\ref{eq:potential-angular-matrix-elements}.
For the transition to $(3,\pm 2, +)$, the 3-j element has a value of $-0.189$, while for $(4, \pm 4, +)$ it is $0.333$.  For this reason, the latter is significantly more favored.

\begin{figure}
    \center
    \includegraphics[scale=0.35]{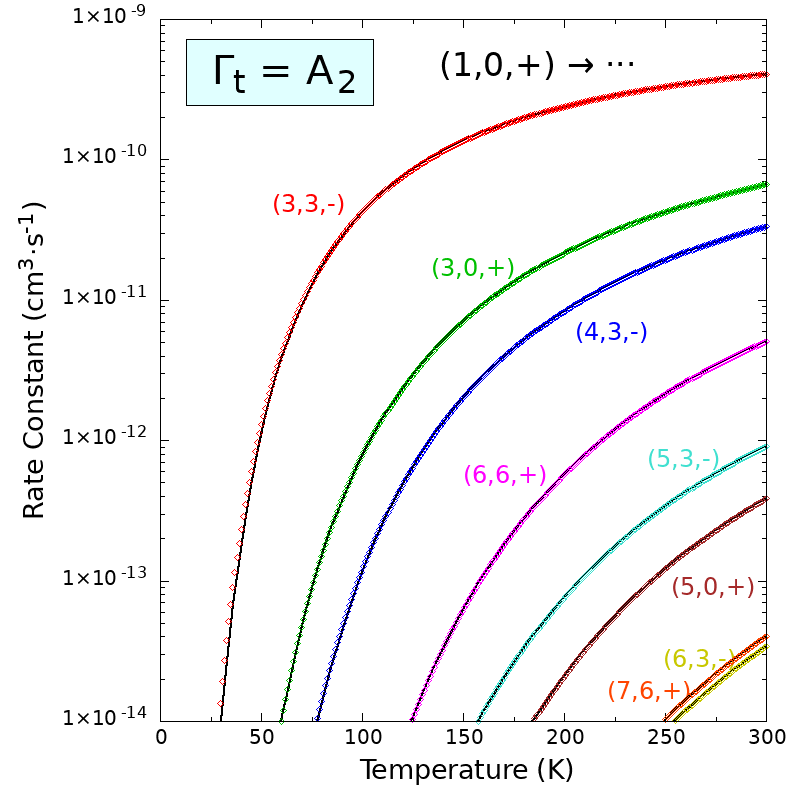}
    
    \includegraphics[scale=0.35]{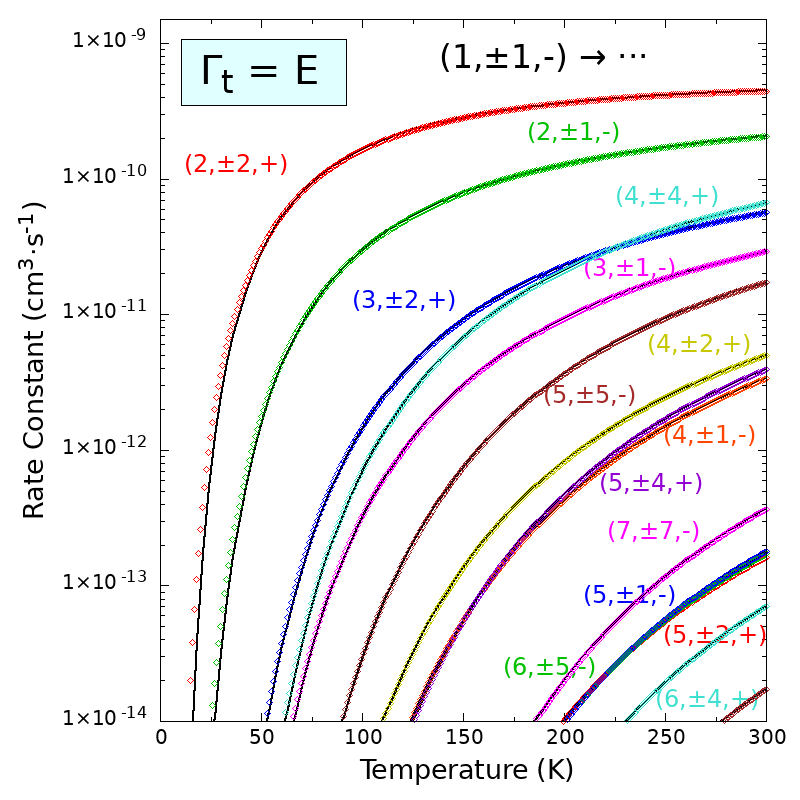}    
    \caption{Calculated state-to-state rate constants for the transitions starting from (1,0,+) within the $\Gamma_t = A_2$ (top panel) and $E$ (bottom panel) representations. The fitted curves are represented by the solid black lines, while the computed data are the coloured dots.}
    \label{fig:A2yE-fit-rate0001.png}
\end{figure}

The rate constants share general tendencies in terms of the relative dominance of the transitions and are shown in Figs.~\ref{fig:A2yE-fit-rate0001.png}, for temperatures between 10 and 300 K, as representative results corresponding to the endothermic transitions starting from (1,0,+) and $(1,\pm 1, -)$ within the $\Gamma_t = A_2$ and $\Gamma_t = E$ representations.

The three parameters $\alpha$, $\beta$ and $\gamma$ obtained from the fit of the
results of Fig.~\ref{fig:A2yE-fit-rate0001.png}  to an Arrhenius-type rate constant 
\begin{equation}
k(T) = \alpha \, \left(\dfrac{T}{300}\right)^\beta \, \exp\left(\dfrac{-\gamma}{T}\right)   
\end{equation}
  are listed in Table~\ref{tab:Parameters-A2yE}, where $k(T)$ is in $\mathrm{cm}^{3}\, \mathrm{s}^{-1}$ and the temperatures are in $\mathrm{K}$.

Additionally, in the S.I. we provide  a file containing the calculated state-to-state rate constants at some selected temperatures, for the de-excitation in this case, in the format required by the Meudon PDR code \citep{2006ApJS..164..506L}. This code computes the steady state of a 1D irradiated slab of interstellar gas by solving for chemical, thermal balance, radiative transfer and statistical equilibrium of many observed species\footnote{Available at \href{https://pdr.obspm.fr}{https://pdr.obspm.fr}}.

\begin{table}
    \centering
    \caption{Fitted parameters for the transitions  ($j,\omega,p_t$) $\rightarrow$ (j',$\omega$',p$_t$' $\Gamma_t$) within $\Gamma_t = A_2$ (1,0,+) and $E$ (1,$\pm$ ,-) representations.}
    \label{tab:Parameters-A2yE}
    \begin{tabular}{c c c c}
    \hline
    \hline
       $(j', \omega', p'_t, A_2)$ & $\alpha \cdot 10^{10} \, (\mathrm{cm}^{3} \, \mathrm{s}^{-1})$ & $\beta$ & $\gamma (K)$ \\ \hline
       (3,3,-) & $14.599$ & -0.26 & 382.66 \\
       (3,0,+) & $5.813$ & 0.12 & 648.88 \\
       (4,3,-) & $6.474$ & -0.22 & 887.95 \\
       (6,6,+) & $5.323$ & -0.36 & 1395.86 \\
       (5,3,-) & $1.659$ & -0.29 & 1560.41 \\
       (5,0,+) & $1.936$ & -0.48 & 1864.97 \\
       (6,3,-) & $0.871$ & -1.02 & 2352.99 \\
       (7,6,+) & $1.004$ & -1.04 & 2345.26 \\
       \hline
       $(j', \omega', p'_t, E)$ & $\alpha \cdot 10^{10} \, (\mathrm{cm}^{3} \, \mathrm{s}^{-1})$ & $\beta$ & $\gamma(K)$ \\
       \hline
       $(2,\pm 2,+)$ & 8.610 & -0.28 & 196.07 \\
       $(2,\pm 1,-)$ & 5.614 & -0.06 & 300.40 \\
       $(3,\pm 2,+)$ & 3.857 & -0.15 & 576.62 \\
       $(3,\pm 1,-)$ & 2.728 & 0.06 & 670.19 \\
       $(4,\pm 4,+)$ & 6.615 & -0.01 & 688.64 \\
       $(5,\pm 5,-)$ & 5.190 & -0.37 & 1020.62 \\
       $(4,\pm 2,+)$ & 1.926 & -0.04 & 1094.03 \\
       $(4,\pm 1,-)$ & 2.01 & 0.06 & 1224.65 \\
       $(5,\pm 4,+)$ & 3.486 & -0.28 & 1344.99 \\ 
       $(5,\pm 2,+)$ & 0.526 & -0.32 & 1732.73 \\
       $(6,\pm 5,-)$ & 0.839 & -0.58 & 1856.99 \\ 
       $(5,\pm 1,-)$ & 0.869 & -0.46 & 1854.82 \\
       $(7,\pm 7,-)$ & 0.220 & -0.70 & 1917.51 \\
       $(6,\pm 4,+)$ & 0.936 & -0.77 & 2157.21 \\
       \hline
    \end{tabular}

\end{table}

\section{Observations and models}\label{sec:Astro}

In Paper I, the authors discuss the excitation of $\mathrm{H}_3^+$ towards $9$ lines of sight (see their Table~2 and references therein). Here, we only consider two of them: HD 110432 and HD 73882. As shown in Table~\ref{tab:Observational-data.}, which summarizes the main observational results, the former has a rather low molecular fraction { defined as  ${2\,n(\mathrm{H}_2) / (2\,n(\mathrm{H}_2) + n(\mathrm{H}))}$} and warm kinetic temperature, while the latter is colder and has a higher molecular fraction. Here, $x_{Obs}(\mathrm{H}_3^+)$ is the ratio of the column densities of $\mathrm{H}_3^+$ and $\mathrm{H}_2$, which is believed to be representative of the ion local fractional abundance, and $f_{Obs}$ is the molecular fraction towards the line of sight derived from data from \citet{2024A&A...685A..82E} as found in Table~1 of \citet{2024arXiv240811511O}. These values use results from the GAIA DR3 release which gives the repartition of absorbing matter along the line of sight towards nearby stars.
The derived molecular fraction is larger than that obtained from the measured column densities of atomic and molecular Hydrogen from ultraviolet observations. That occurrence is due to the possible presence of atomic hydrogen on the line  of sight that does not belong to the considered interstellar translucent cloud. The distances of HD110432 and HD73882 are respectively $438 \, \mathrm{pc}$ and $461 \, \mathrm{pc}$ (obtained from their measured parallax available on the SIMBAD database), the sizes of the clouds are about a few $\mathrm{pc}$, so that even a very small atomic hydrogen volumic density, located randomly on the line of sight, may contribute to the atomic column density by a non negligible amount.

\begin{table*}[ht]
\caption{Observational data.\protect\label{tab:Observational-data.}}
\medskip{}

\centering%
\begin{tabular}{c c c c c c c c c}
\hline
\hline
\multirow{2}{*}{HD} & $E_{\mathrm{B-V}}$ & $A_{\mathrm{V}}$ & $N\left(\mathrm{H}_{3}^{+}\right)$ & $N\left(\mathrm{H}_{2}\right)$ &  $x_{Obs}\left(\mathrm{H}_{3}^{+}\right)$ & $T_{01}(\mathrm{H_{2}})$ & $T_{12}(\mathrm{H}_{3}^{+})$ & $f_{Obs}$\tabularnewline
 & $\mathrm{mag}$ & $\mathrm{mag}$ & $10^{13}\,\mathrm{cm}^{-2}$ & $10^{20}\,\mathrm{cm}^{-2}$ & $\times10^{-8}$ & $\mathrm{K}$ & $\mathrm{K}$ & \tabularnewline
\hline
110432 & $0.40$ & $2.02\pm0.33$ & $5.2$ & $4.4$ & $11.8$ & $68$ & $30$ & $0.56$\tabularnewline
73882 & $0.72$ & $2.36\pm0.23$ & $9.0$ & $13.1$ & $6.87$ & $51$ & $23$ & $0.85$ \tabularnewline
\hline
\end{tabular}
\end{table*}

Using the gas temperature deduced from $\mathrm{H}_2$ excitation and the observed molecular fraction, we can use the code \texttt{ExcitH3+} \footnote{Available at \href{https://excith3p.ism.obspm.fr}{https://excith3p.ism.obspm.fr}} to test the impact of various rate coefficients on $\mathrm{H}_3^+$ excitation.

\begin{figure}
    \center
    \includegraphics[width=1.0\columnwidth]{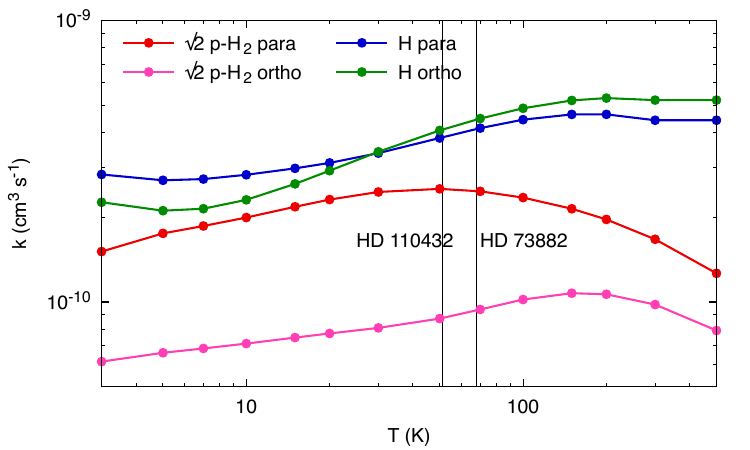} 
    \caption{Comparison of new $\mathrm{H}$ collision rate coefficients with the proxy used in \citet{2024MolPh.12282612L} for the lowest {para} and {ortho} transitions. The temperatures of HD 110432 and HD 73882, as derived from molecular hydrogen $J=1$ and $J=0$ column density ratio $T_{01}(\mathrm{H}_2)$, are indicated by a black vertical line.}
    \label{fig:Cmp_H3p.png}
\end{figure}

We compare the impact of the new rates of collision to the proxy used in Paper I: $k_{\mathrm{H}} = \sqrt{2} \, k_{\mathrm{pH}_2}$, where $k_{\mathrm{pH}_2}$ are the rate coefficients of $\mathrm{H}_3^+$ with {para} $\mathrm{H}_2$, taken from \citet{Gomez-Carrasco-etal:12}. Figure~\ref{fig:Cmp_H3p.png} shows the rate coefficients for transitions between levels $(2,2,+)$ and $(1,1,-)$ (lowest {para}) and $(3,3,-)$ and $(1,0,+)$ (lowest {ortho}). We see that the new rates are stronger than the proxy, and that the values involving {ortho} transitions on one hand and {para} transitions on the other are much closer. Collisions with electrons are included, using the results of \citet{Kokoouline:2010}.

Paper I emphasized the important role of $\mathrm{H}_3^+$ destruction rate $k_{DR}$ that is due to dissociative recombination,
for which the values pertaining to the {para} form, $k_{DR}^p$, and to the {ortho} form, $k_{DR}^o$, are introduced. 
As in Paper I, we adopt a standard electronic recombination rate of $\mathrm{H}_3^+$
with electrons of $k_{DR}^p(T) = 5.23 \, 10^{-8} \, \left( \frac{T}{300}\right)^{-0.75} \, \mathrm{cm}^3\,\mathrm{s}^{-1}$. However,
Paper I stresses that the exact rate is still not well known and,
in particular, that {ortho} and {para} levels may have different rates.
\citet{Pagani-etal:09} suggests a constant rate of $k_{DR}^o = 6 \, 10^{-8} \, \mathrm{cm}^3\,\mathrm{s}^{-1}$ below $250 \, \mathrm{K}$ for {ortho}-$\mathrm{H}_3^+$ (see discussion in Section 3.3 of Paper I).
So, for each line of sight we use three values of the ratio $k_{DR}^o / k_{DR}^p$ (close to $1/3$, $2/3$ and $1$),
corresponding respectively to the rates of \citet{Pagani-etal:09} {  at 60~K}, an intermediate case and identical rate coefficients.

\begin{figure}
\centering\includegraphics[width=1.0\columnwidth]{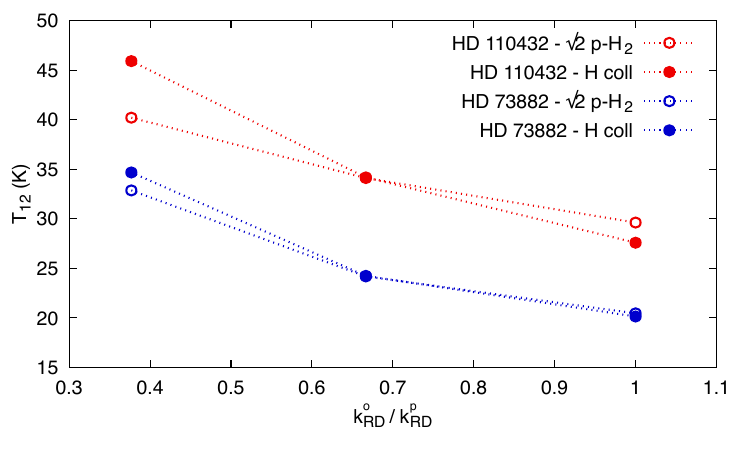}
\caption{Computed values of $T_{12}$ for the two lines of sight and three ratios of $k_D^o / k_D^p$. Proxy: $k_\mathrm{H} = \sqrt{2} \, k_{\mathrm{pH}_2}$, H coll: $k_\mathrm{H}$ from this work.\protect\label{fig:Values-of-T12}}
\end{figure}

This leads to $6$ different estimates of $T_{12}(\mathrm{H}_3^+)$ for each line of sight
($3$ recombination rates for $\mathrm{H}_3^+$, times $2$ sets of collision rate coefficients with $\mathrm{H}$).
They are presented on Figure~\ref{fig:Values-of-T12}.

We see that changing the ratio $k_{DR}^o / k_{DR}^p$ has a strong impact on the computed value of $T_{12}(\mathrm{H}_3^+)$. In addition, using the {para}-$\mathrm{H}_2$ proxy leads to a reduced amplitude of the variations of $T_{12}(\mathrm{H}_3^+)$ with $k_{DR}^o$. A particularly remarkable result is that, for a given line of sight, the $T_{12}(\mathrm{H}_3^+)$ curves cross for an intermediate value of $k_{DR}^o / k_{DR}^p$. So, for that specific value only, there is no difference in the excitation temperature between the two sets of collision rate coefficients. As expected, the impact of the new rates is larger for lower values of the molecular fraction, as in HD 110432 compared to HD 73882.

Studying the origin of the sensitivity of $T_{12}(\mathrm{H}_3^+)$ to $k_{DR}^o / k_{DR}^p$ is out of the scope of this work, and will be done in another paper. However, regardless of the final determination of the recombination rates, it remains essential to use the new collision rate coefficients with $\mathrm{H}$, as they result in significantly different evaluations of $T_{12}(\mathrm{H}_3^+)$. Depending on the line of sight considered and the choice of other physical parameters, the differences may reach up to $20 \, \%$ compared to using the proxy.

\section{Conclusions}

In this paper we introduce new accurate computations of collisional cross sections and collisional rate coefficients of $\mathrm{H}_3^+$ with $\mathrm{H}$ for energies adapted to the physical conditions in the Interstellar Medium.

A time independent close-coupling method is used to calculate the state-to-state cross sections and rate coefficients, using a very accurate and full dimensional potential energy surface recently developed for this system \citep{delMazo-Sevillano-etal:24a}. H$_3^+$ rovibrational levels are calculated using hyperspherical coordinates, in permutationally symmetry adapted functions up to $j=20$ and including the vibrations.
From all these levels, only those corresponding to the ground (0,0$^0$)
vibrational state are selected and included in the close-coupling method
in a symmetric top approach, considering  frozen H$_3^+$ as equilateral triangle but with the numerically exact energy levels obtained in full-dimensional H$_3^+$ rovibrational states.

We use these new rate coefficients to evaluate the excitation temperature of $\mathrm{H}_3^+$ as determined from its $2$ lowest levels.
Compared to using a proxy based on {para} $\mathrm{H}_2$, we find that:
\begin{itemize}
    \item Differences in the estimated excitation temperature may reach $20 \%$.
    \item The impact of the new rate coefficients increases as the cloud molecular fraction decreases.
    \item Differences between electronic recombination rates of {ortho} and {para} $\mathrm{H}_3^+$
      have a significant impact on the estimated excitation temperature, with differences larger with the new $\mathrm{H}$ collision rates.
\end{itemize}

This later point leads to large difficulties in estimating physical conditions from the observed values of $\mathrm{H}_3^+$ abundance and excitation.
The reasons for this sensitivity will be explored in a future paper, but it stresses the necessity to still improve our understanding of this long standing question.

In the present work, no estimation is done of possible reactive collisions between $\mathrm{H}$ and $\mathrm{H}_3^+$
that may lead to {ortho} to {para} transitions.
Inspection of the PES shows that these rates must be low (lower than $10^{-12} \, \mathrm{cm}^3 \, \mathrm{s}^{-1}$ { below 100 K}), but even a low value may have an impact in some specific conditions, and that question too deserves further examination.

As a last point, the present work does not include radiative pumping in the infrared range, which is invoked in \citet{Pereira-Santaella-etal:24} to explain their observations of vibrationally excited $\mathrm{H}_3^+$ in (U)LIRGs.

\section{ Data availability}

The energy levels of H$_3^+$ obtained in this work are listed  in file \href{https://nestor.aanda.org/aa/user/article/aa52977-24}{H3p-levels-exomol.dat}.

The inelastic rates of ortho  and para symmetries are provided in the files
\href{https://nestor.aanda.org/aa/user/article/aa52977-24}{A2-rates.dat} and
\href{https://nestor.aanda.org/aa/user/article/aa52977-24}{E-rates.dat} files, respectively. 

\begin{acknowledgements}
 The research leading to these results has received funding from
 MICIN (Spain), under grant PID2021-122549NB-C21 and
 PID2021-122549NB-C22, and by the Programme National « Physique et Chimie du Milieu Interstellaire » (PCMI) of the CNRS/INSU with INC/INP co-funded by CEA and CNES.
 Computational assistance was provided by the Supercomputer facilities of Lusitania founded by the C\'enitS and Computaex Foundation.
\end{acknowledgements}

\bibliographystyle{aa}

\clearpage{}

\begin{appendix}
\section{H$_3^+$ levels\label{sec:H3+-levels}}

For the $\mathrm{H}_3^+$ we use the PES for $\mathrm{H}_4^+$ of \citet{delMazo-Sevillano-etal:24a}, placing the fourth hydrogen atom at $100 \, \mathrm{bohr}$ at a fixed geometry. The rovibrational bound states of $\mathrm{H}_3^+$ are calculated in the adiabatically adjusting, Johnson's principal axis, hyperspherical (denoted APJH)  coordinates \citep{Pack-Parker:89} using a code previously developed \citep{Aguado-etal:00,Sanz-etal:01}, in which the full wave function is expanded as
\begin{eqnarray}\label{eq:total-function-h3+}
    && \Psi^{jm\Gamma}_{i} (\rho,\theta,\phi_\tau,\alpha,\beta,\gamma)
    = 4\rho^{-5/2}\,\sum_{v,k,n,\Omega} C^{jm\Gamma i}_{v,k,n,\Omega} \nonumber\\ 
    &\times& \tilde{W}^{jm\Gamma}_{\Omega,n}(\alpha,\beta,\gamma,\phi_\tau)
    \, F^{j,\Omega,n}_k(\theta) \, \varphi_v(\rho), 
\end{eqnarray}
where, $\varphi_v(\rho)$ are numerical functions obtained in a mono-dimensional radial grid in $\rho$ for the equilibrium configuration \citep{Aguado-etal:00},
$F^{j,\Omega,n}_k(\theta)$ are proportional to Jacobi polynomials \citep{Aguado-etal:00,Abramowitz-Segun:72} and $ \tilde{W}^{jm\Gamma}_{\Omega,n}(\alpha,\beta,\gamma,\phi_\tau)$ are symmetry adapted basis set functions for a given triatomic angular momentum $j$, with projection $m$ on the space fixed frame and an irreducible representation, $\Gamma$, of the D$_{3h}$ point group (isomorphic with the $S_3$ permutation group multiplied by the inversion of spatial symmetry). These symmetry adapted functions are expressed as
\begin{eqnarray}\label{d3h-functions}
     \tilde{W}^{jm\Gamma}_{\Omega,n}(\alpha,\beta,\gamma,\phi_\tau)
     &=& A^{j\Gamma}_{\Omega n} \quad W^{jm}_{\Omega,n}(\alpha,\beta,\gamma,\phi_\tau)\\
     &+& B^{j\Gamma}_{\Omega n} \quad W^{jm}_{-\Omega,-n}(\alpha,\beta,\gamma,\phi_\tau),\nonumber
\end{eqnarray}
where $A^{j\Gamma}_{\Omega n}$ and $B^{j\Gamma}_{\Omega n}$
coefficients are obtained applying projection operators \citep{Aguado-etal:00,Sanz-etal:01}, and
\begin{equation}\label{angular-functions}
 {W}^{jm}_{n\omega} (\alpha,\beta,\gamma,\phi_\tau) =
 \sqrt{\dfrac{2j+1}{8\pi^2}} D^{j*}_{m\omega}(\alpha,\beta,\gamma) 
 \dfrac{e^{i n \phi_\tau}}{\sqrt{2\pi}},
\end{equation}
with $D^{j*}_{m\omega}(\alpha,\beta,\gamma)$ being Wigner rotation functions \citep{Zare-book}, transforming from the space fixed to the body-fixed frame, with the axes along the principal inertia axes and the z-axis perpendicular to the plane of $\mathrm{H}_3^+$, with $\omega$ being the projection of the triatomic total angular momentum in the body-fixed z-axis.

The exact triatomic eigenvalues and eigenvectors are obtained for each $j,\Gamma$ values using a two steps method \citep{Aguado-etal:00}. The eigenvalues are obtained using a non-orthogonal Lanczos method \citep{Cullum-Willoughby:85} and then the eigenfunctions are obtained using a conjugate gradient method \citep{Froberg}. From the eigenvectors the approximated quantum numbers ($v_1,v_2^\ell$) \citep{Watson:84} are obtained together with the $\omega$ distribution in Eq.~\ref{eq:total-function-h3+}.
Bound state calculations have been done up to $j=15$ for  {ortho}-$\mathrm{H}_3^+$  (nuclear spin $I=3/2$ and $\Gamma= A_2',A_2''$) and {para}-$\mathrm{H}_3^+$ ($I=1/2$, $\Gamma=  E' , E''$), which are listed in the Supplementary Information (S.I.)  with the  quantum numbers.

\subsection{Simplifying to  rigid rotor $\mathrm{H}_3^+$}
The scattering is treated in the rigid rotor approach, and not all the bound states are used. 
 In what follows, we shall consider $\Gamma = \Gamma_t \times p_t$, to treat separately the permutation symmetry, $\Gamma_t$ which is conserved in inelastic collisions, and the inversion of spacial coordinates, $p_t$, which is not conserved
 ($p_t$=1 for $A_2'$ and $E'$, $p_t$= -1  for $A_2''$ and $E''$).
 To include the "exact" triatomic bound states in the scattering calculations described below, the following approximations have been done:
\begin{enumerate}

 \item Only states corresponding to the ground vibrational level,  $(0,0^0)$, are included, up
 to an energy of $4100 \, \mathrm{cm}^{-1}$ above the ground state.

 \item Triatomic systems are in general asymmetric top. However, for the triangular equilibrium distance, $\mathrm{H}_3^+$ is a symmetric top. In the exact  bound calculations, the states in general are a linear superposition of several $\omega$ values, but for $(0,0^0)$ and low triatomic angular momenta
 $j$ there is a dominant $\omega$ value, with a weight larger than 90\% (see S.I.). Thus, in the present case we shall consider $\mathrm{H}_3^+$ as a symmetric top in the collisions studied below. 

 \item Under this symmetric top rigid-rotor approximation, bound states of Eq.~\ref{eq:total-function-h3+} are taken as
 \begin{eqnarray}\label{eq:rigid-rotor-function-H3+}
     \left\vert j m \omega p_t \Gamma_t\right\rangle \equiv  \sqrt{{2 j+1 \over 8\pi^2}} \, D^{j*}_{m\omega}(\alpha,\beta,\gamma),
 \end{eqnarray}
with the only restriction of adding an intrinsic parity under inversion of spatial coordinates, 
so that the parity of these functions coincide with the corresponding parity of the exact triatomic state, $p_t$. The functions thus defined have the proper permutation symmetry, $A_2$ or $E$ in the present case.

 \item For $\Gamma=E'$ or $E''$, the functions $\tilde{W}^{jm\Gamma}_{n\omega}
 = f\Big({W}^{jm\Gamma}_{n\omega}, {W}^{jm\Gamma}_{-n-\omega}\Big)$ are decoupled from $\tilde{W}^{jm\Gamma}_{n-\omega}
 = f\Big({W}^{jm\Gamma}_{n-\omega}, {W}^{jm\Gamma}_{-n\omega}\Big)$, with $n,\omega$ different from zero. Thus, the two degenerate eigenstates are expressed in the corresponding basis functions separately, and labeled by $\omega$ and $-\omega$. 
\end{enumerate}

\section{Inelastic scattering method \label{sec:symmetric-top-scattering}}

The inelastic scattering calculations have been performed with the code DTICC which has been specifically developed by the authors for this purpose, in which the close coupled equations are solved with a renormalized Numerov algorithm \citep{Gadea-etal:97}. The system composed by a symmetric top colliding with an atom is described in space fixed (SF) coordinates, using a  standard method adapted from $\mathrm{NH}_3$ \citep{Green1980}, outlined here for completeness.

\begin{figure}
    \centering
    \includegraphics[width=0.75\linewidth]{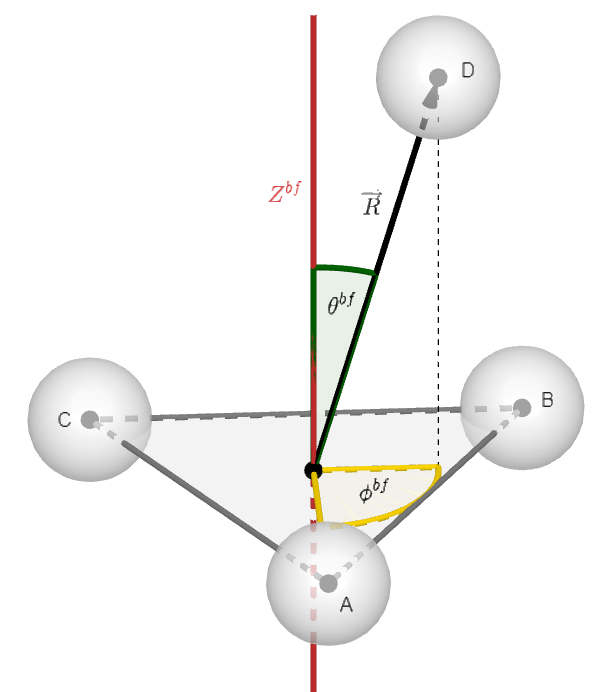}
    \caption{Relative orientation between both fragments, in a { body-fixed} frame centered in the triatom.}
    \label{fig:H4p-BF}
\end{figure}

A Jacobi vector  $\vec{R} \equiv \{ R, \theta_R, \phi_R\}$ is defined in the SF frame joining the ABC (or symmetric top) center-of mass to the colliding atom D, whose associated angular momentum is   $\ell$. The
space-fixed angular basis set functions are defined as 
\begin{equation}
 \ket{\alpha} =
 \sum_{m m_{\ell}} \bra{j m \ell m_\ell}\ket{JM} \ket{j m \omega p_t \Gamma_t} 
 Y_{\ell m_{\ell}}(\theta_R, \phi_R),
\end{equation}
with  $\ket{j m \omega p_t \Gamma_t} $ being defined in Eq.~\ref{eq:rigid-rotor-function-H3+}. The $\ket{\alpha}= {\cal Y}^{JMp}_{j\omega\ell p_t \Gamma_t}(\alpha,\beta,\gamma,\theta_R,\phi_R)$ is introduced to simplify the notation.
In this basis, the total scattering wave function is expressed as
\begin{eqnarray}\label{eq:total-scattering-function}
    \ket{\Psi^{JM\alpha-}_E} = \sum_{\alpha'} {\Phi^{JM\alpha-}_{\alpha'}(R;E)\over R} \ket{\alpha'}
\end{eqnarray}
where the radial coefficients $\Phi^{JM\alpha-}_{\alpha'}(R;E)$ are obtained numerically by solving the close coupled differential equations obtained by inserting this total wave function, with the Hamiltonian being defined as

\begin{equation}
\begin{split}
 \hat{H} = &-\dfrac{\hbar^2}{2\mu}
 \left( \dfrac{2}{R}\dfrac{\partial}{\partial R} + \dfrac{\partial^2}{\partial R^2} \right)
 + \dfrac{\hat{\ell}^2}{2\mu R^2} \\
& + V(R,\theta^{bf},\phi^{bf}) 
 + \hat{H}_{ABC}
\end{split}
\end{equation}
where $\mu=m_D m_{ABC}/(m_D+m_{ABC})$ is the reduced mass, $\hat{H}_{ABC}$ is the internal triatomic hamiltonian operator, and $V(R,\theta^{bf},\phi^{bf})$ is the interaction potential which only depends on the relative orientation of the atom D with respect to the triatomic fragment ABC --- as shown in Fig. \ref{fig:H4p-BF}.

\begin{figure}
    \centering
    \includegraphics[width=8cm]{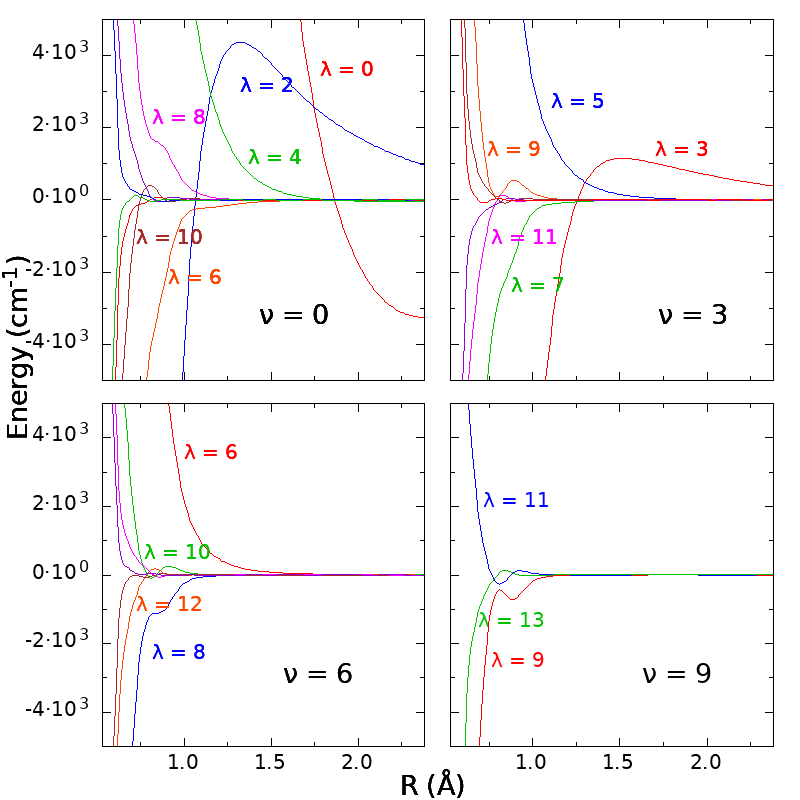}
    \caption{Radial coefficients corresponding to the expansion of the interaction potential.}
    \label{fig:Vlamnu}
\end{figure}

The centrifugal term $\hat{\ell}^2$ is diagonal in the SF representation with eigenvalues $\hbar^2\,\ell(\ell+1)$.
The eigenstates and eigenvalues of the ABC system were obtained in the previous section, 
which in the present treatment are simplified to the symmetry adapted symmetric top functions, $\ket{jm\omega p_t \Gamma_t}$.
The potential matrix elements in the space-fixed representation are then given by
\begin{eqnarray}
    \bra{\alpha} V \ket{\alpha'}= \sum_{\lambda \nu} V_{\lambda \nu} (R) 
     \bra{\alpha} Y_{\lambda\nu} \ket{\alpha'},
\end{eqnarray} 
where $V_{\lambda \nu} (R)$ are the radial coefficients of the expansion of the
  potential in spherical harmonics as
\begin{equation}\label{eq:potential-expansion}
 V(R,\theta^{bf},\phi^{bf}) = \sum_{\lambda \nu} V_{\lambda \nu} (R) 
 Y_{\lambda \nu}(\theta^{bf},\phi^{bf}),
\end{equation}
which are shown in Fig.~\ref{fig:Vlamnu}, and
\begin{equation}\label{eq:potential-angular-matrix-elements}
\begin{split}
&\bra{\alpha}Y_{\lambda\nu}\ket{\alpha'} =
\sqrt{\dfrac{[\ell][\lambda][\ell'][j][j']}{4\pi}} 
(-1)^{-J+\lambda-\omega'} \\
&\left(\begin{array}{c c c}
\ell & \lambda & \ell' \\
0 & 0 & 0
\end{array}\right)
\left\{\begin{array}{c c c}
j & J & \ell \\
\ell' & \lambda & j'
\end{array}\right\}
\left(\begin{array}{c c c}
j & \lambda & j' \\
\omega & \nu & -\omega'
\end{array}\right),
\end{split}
\end{equation}
where the compact notation $[j] = 2j + 1$ has been used for the angular momenta degeneracies, and the 6-j symbols \citep{Zare-book} are denoted as $\{ :~:~:\}$.  The symmetry of the potential ensures that the radial expansion coefficients fulfill the condition $V_{\lambda\nu}(R) = (-1)^\nu V_{\lambda -\nu}(R)$, as well as the threefold symmetry for the azimuthal angle $ V(R,\theta^{bf},\phi^{bf}) =  V(R,\theta^{bf},\phi^{bf} + {2n\pi}/{3})$ for $n \in \mathbb{Z}$.
These symmetry constraints are equivalent to say that the only allowed values for $\nu$ are multiples of three: $\nu = 0, \pm 3, \pm 6, \dots$.
The total parity of the four atom system is conserved and equal to $p=p_t (-1)^\ell$, so that $p_t$ = $\pm 1$ is no longer conserved.

As noted above, the symmetry adapted rigid rotor wave functions of $\mathrm{H}_3^+$ used in this work constitute
a slight modification, essentially equivalent, of the usual treatment for collisions of symmetric top functions with atoms \citep{Green1976,Green1980}. First, the procedure followed allows to exactly determine the symmetry of the triatomic function, which also depends on the hyperspherical angle $\phi_\tau$, which is not included in the rigid rotor approach. 

The state-to-state cross section are obtained using the usual partial wave summation in the space fixed frame \citep{Arthurs1960} as
\begin{equation}\label{eq:s2s-cross-section}
\sigma^{\Gamma_t}_{\beta\beta'}(E) = \dfrac{\pi}{(2j+1)k_\beta^2} 
\sum_{J\ell\ell' p}(2J+1) {\vert S^{J\Gamma_t p}_{\alpha\alpha'} (E) - \delta_{\alpha\alpha'}\vert}^2
\end{equation}
where $k_\beta^2 = {2\mu (E- E_\beta)}/{\hbar^2}$, $E$ is the total energy, and $E_\beta$ is the energy of the triatomic level $\beta$. $S^{J \Gamma_t p}_{\alpha\alpha'} (E)$ are the elements of the scattering matrices obtained  in the resolution of the close coupling equations.

The state-to-state rate constants are obtained by taking Boltzmann averages on the { cross-sections} --- {\it i.e.} integration over the kinetic energy $E^k_\beta = E - E_\beta$ for a given temperature $T$, as
\begin{equation}
\begin{split}
 &k_{\beta\beta'}(T) = \sqrt{\dfrac{8}{\pi\mu(k_B T)^3}} ~\times\\
&\times~\int dE^k_\beta ~E^k_\beta ~\sigma_{\beta\beta'}(E^k_\beta) 
~e^{-\frac{E^k_\beta}{k_B T}}   
\end{split}
\end{equation}
being $k_B$ the Boltzmann constant.

\end{appendix}

\end{document}